\documentclass[useAMS,usenatbib,onecolumn,preprint]{mn2e}

\usepackage{txfonts}
\usepackage{graphicx}

\def\microK{\mu\mathrm{K}}

\def\Hwedge{H_{\mathrm{KBOE}}}

\def\XWmean{\bar{I}_{\mathrm{KBOE}}}

\def\AW{A_{\mathrm{KBOE}}}

\def\atzwedge{a_{2,0}^{\mathrm{KBOE}}}

\def\ndust{n}
\def\Mdust{\mathcal{M}_{\mathrm{dust}}}
\def\Rin{R_{\mathrm{in}}}
\def\Rout{R_{\mathrm{out}}}
\def\dustsize{a}

\def\Tdust{T}
\def\taudust{\tau}
\def\IKBOE{I_{\lambda}^{\mathrm{KBOE}}}
\def\IZLE{I^{\mathrm{ZLE}}_{\lambda}}
\def\ITOT{I^{\mathrm{TOT}}_{\lambda}}

\def\Cl{C_{\ell}}

\def\lsim{\,\lower2truept\hbox{${<\atop\hbox{\raise4truept\hbox{$\sim$}}}$}\,}
\def\gsim{\,\lower2truept\hbox{${>\atop\hbox{\raise4truept\hbox{$\sim$}}}$}\,}

\def\Planck{{\sc Planck}}

\def\Point{\mbox{$\hat{P}$}}

\def\var{\mbox{$\mathrm{var}$}}

\def\Nside{\mbox{$N_{\mathrm{side}}$}}

\title[
Large Scale Traces 
of Solar System Cold Dust on CMB Anisotropies
]{
Large Scale Traces 
of Solar System Cold Dust on CMB Anisotropies
}
\author[M. Maris, C. Burigana, A. Gruppuso, F. Finelli and J.M. Diego  ]{M. Maris$^{1}$\thanks{E-mail:
maris@oats.inaf.it}, C. Burigana$^{2}$, A. Gruppuso$^{2,4}$, 
F. Finelli$^{2,4}$ and J.M. Diego $^{3}$\\
$^{1}$INAF - Osservatorio Astronomico di Trieste, Via G.B.Tiepolo 11, I34100, Trieste, Italy\\
$^{2}$INAF - IASF Bologna, Via P. Gobetti 101, Bologna, I40129, Italy\\
$^{3}$IFCA, Instituto de Fisica de Cantabria (UC-CSIC). Avda. Los Castros s/n. 39005 Santander, Spain\\
$^{4}$INFN, Sezione di Bologna, Via Irnerio 46, I-40126 Bologna, Italy}
\begin{document}

\date{Accepted 2011 April 08. Received 2011 April 08; in original form 2010 October 04}

\pagerange{\pageref{firstpage}--\pageref{lastpage}} \pubyear{2011}

\maketitle

\label{firstpage}

\begin{abstract}
We explore the microwave anisotropies at large angular scales produced by 
the emission from cold and large dust grains, expected to exist in the outer parts of the Solar System,
using a simple toy model for this diffuse emission. 
Its amplitude is constrained in the Far--IR
by the COBE data and is compatible with simulations found in the literature.
We analyze the templates derived after subtracting our model  
from the WMAP ILC 7~yr maps and investigate on the cosmological implications of such 
a possible foreground. The anomalies related to the low quadrupole 
of the angular power spectrum, the two-point correlation function, the parity 
and the excess of signal found in the ecliptic plane are significantly alleviated.
An impact of this foreground for some cosmological parameters 
characterizing the spectrum of primordial density perturbations, 
relevant for on-going and future CMB anisotropy experiments, is found. 
\end{abstract}

\begin{keywords}
	Interplanetary Medium
--
	(Cosmology): Cosmic Microwave Background
--
	Infrared: Solar System
--
	Submillimeter 
\end{keywords}

\section{Introduction and toy model}\label{sec:introduction}

The thermal emission from Interplanetary Dust Particles (IDPs), 
called Zodiacal Light Emission (ZLE), is a well known source
of foreground in the (Far)IR sky. 
Recently \citet{Maris:etal:2006} discussed the use of the 
ZLE COBE model 
	\citep{Kelsall:etal:1998,Fixsen:Dwek:2002} 
below 1~THz to derive  
predictions for surveys
dedicated to Cosmic Microwave Background (CMB),
such as WMAP \footnote{http://lambda.gsfc.nasa.gov/} and \Planck  \footnote{http://www.rssd.esa.int/planck}.
A weak, albeit significant, contribution
at frequencies $\lsim 857$~GHz will be likely detectable
by \Planck. The simple scaling law of
\citet{Fixsen:Dwek:2002} predicts a small but not negligible contribution at 
70--150~GHz, of particular interest in CMB studies.
Theoretical estimates \citep{Babich:etal:2007,Babich:Loeb:2007} suggest that the collective
emission of Kuiper Belt Objects (KBOs)
and other minor bodies could produce imprints in CMB surveys.

CMB anisotropy maps derived from WMAP data are well consistent 
with the standard $\Lambda$CDM cosmological model.
However, there are some intriguing deviations at large angular scales, such as  
those related to the angular power spectrum (APS)
at low multipoles $\ell$'s
\citep{Bennet:etal:2003a,Hinshaw:etal:2003,Spergel:etal:2003,Larson:2010gs,Kim:2010gd,Kim:2010gf,Gruppuso:2010nd}.
% \citep{Spergel:etal:2006,Huterer:2006}
Other anomalies regard 
the alignment of low multipoles
\citep{Tegmark:etal:2003,Gordon:etal:2005,Copi:etal:2006,Huterer:2006,Copi:etal:2006,Huterer:2006,Gruppuso:2009ee,Gruppuso:2010up}
and to the North--South asymmetry of the 
APS with respect to the ecliptic plane  \citep{Tegmark:etal:2003,Oliveira:etal:2004,Copi:etal:2006,Huterer:2006}.
Moreover, \citet{Vielva04} detected a localized non-Gaussian behavior in the southern
hemisphere using a wavelet analysis technique, see also \citep{Cruz05}, and 
\cite{Diego:etal:2009} recently identified an intriguing excess of emission in 
WMAP 5 years ILC map along the ecliptic plane. 
Their results will be discussed later in more detail.

If not produced by systematic effects 
\citep{Tegmark:etal:2003,Huterer:2006,Burigana:etal:2006,Gruppuso:etal:2007}
these features could be of cosmological origin 
\citep{Gordon:etal:2005,Ghosh:etal:2007}
or effects produced by
some foreground within the
near Universe \citep{Vale:2005,Cooray:Seto:2005,Inoue:Silk:2006a,Inoue:Silk:2006b,Rakic:etal:2006},
the Galaxy \citep{Frisch:2005},
or the Solar System \citep{Schwarz:etal:2004,Copi:etal:2006}.

It can not be excluded 
that some diffuse Solar System emission 
could contribute at low $\ell$'s at WMAP frequencies
\citep{Schwarz:etal:2004,Copi:etal:2006}. 
Such foreground would exhibit planar symmetry with respect to the ecliptic (or 
some plane slightly tilted with respect to it).
According to (Far)IR measures, the ``classical'' 
%or ``standard'' 
ZLE is the thermal 
emission of sub-mm IDPs, with orbital radii within 5~AU, characterized by a $\approx\nu^4$ 
scaling below 1~THz \citep{Fixsen:Dwek:2002}. 
If larger particles exist, their emission
would add to the classical ZLE, producing a
signal at mm wavelengths,
ranging between about 
$1/10$ and $10$ times the classical ZLE,
with a less steep spectrum, 
$\approx1/\nu$,
as suggested by several models
\citep{Backman:etal:1995,Stern:1996,Yamamamoto:Mukai:1998,Moro:2002,Moro:2003}. 
Such particles can be produced by the 
erosion of KBOs due to mutual collisions or by the erosion of incoming
interstellar dust and 
they should be located beyond Jupiter. They would be then colder and possibly larger than the 
IDPs responsible for the classical ZLE.
Impact detectors onboard interplanetary probes reveal that such particles 
indeed exist \citep{Landgraf:etal:2002}, but very little is known about their nature.

Motivated by these considerations, we reconsider the problem of diffuse Solar System emission at cosmological
frequencies, by introducing a toy model of emission from the  KBO dust particles, 
that we will call in this paper KBO Emission (KBOE), and evaluating its implications for CMB 
observations.

%{\em to be modified\\
%We produce and characterize a template map of the ZLE as seen from WMAP 
%which models the expected contribution of ZLE in WMAP frequency channels after averaging 
%7--years of observations, by
%simulating the WMAP scanning strategy and by using a 
%toy ``toy'' model useful to catch many of 
%the features of the map and we discuss briefly the ``seasonal'' modulation lost
%in the averaging process.
%}

\begin{figure}
\centering
 \begin{tabular}{c}
%  \vskip {-0.8cm}
{\bf a)}
{\includegraphics[width=120mm,angle=0]{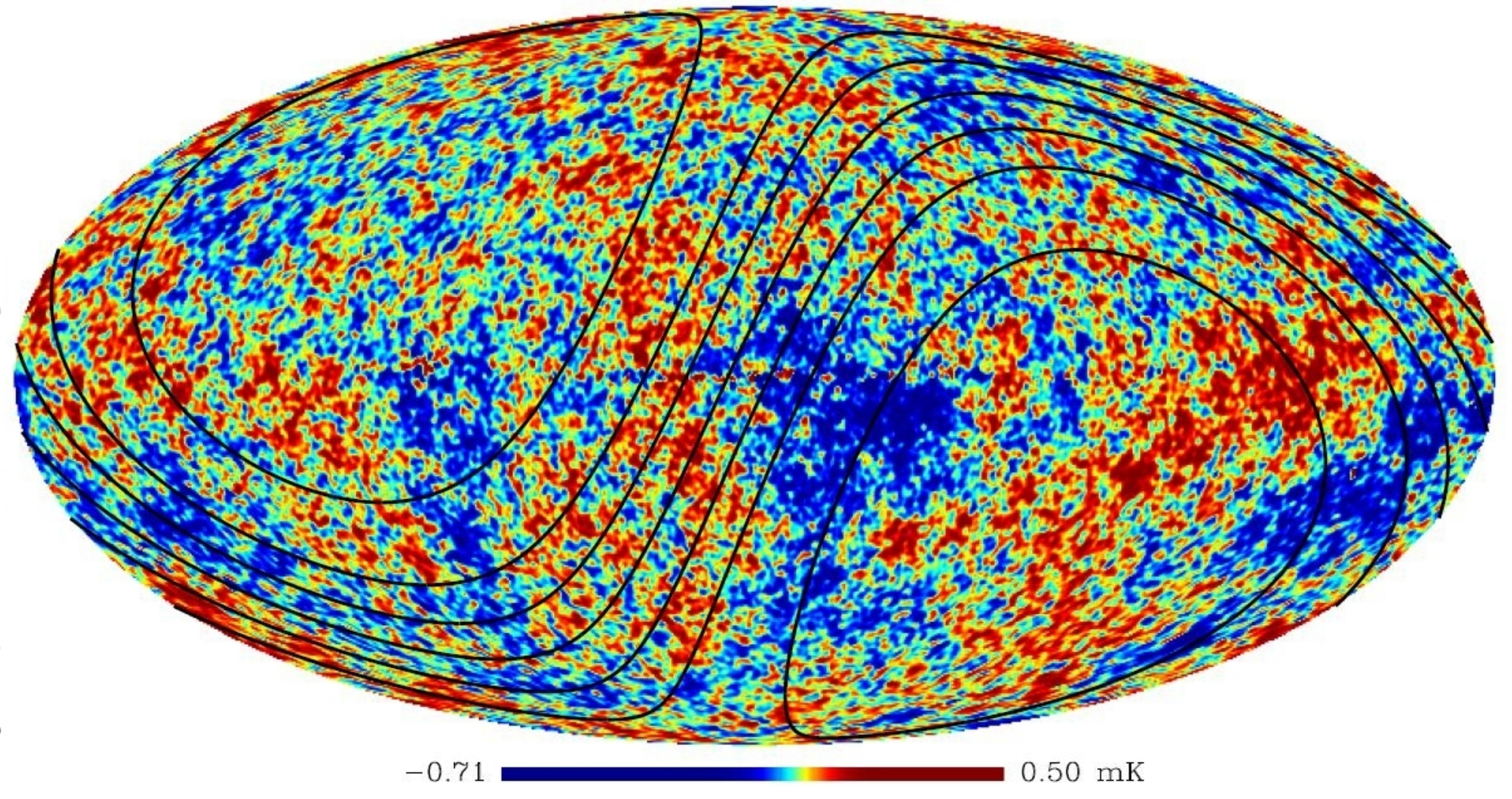}} \\
{\bf b)}
 {\includegraphics[width=140mm]{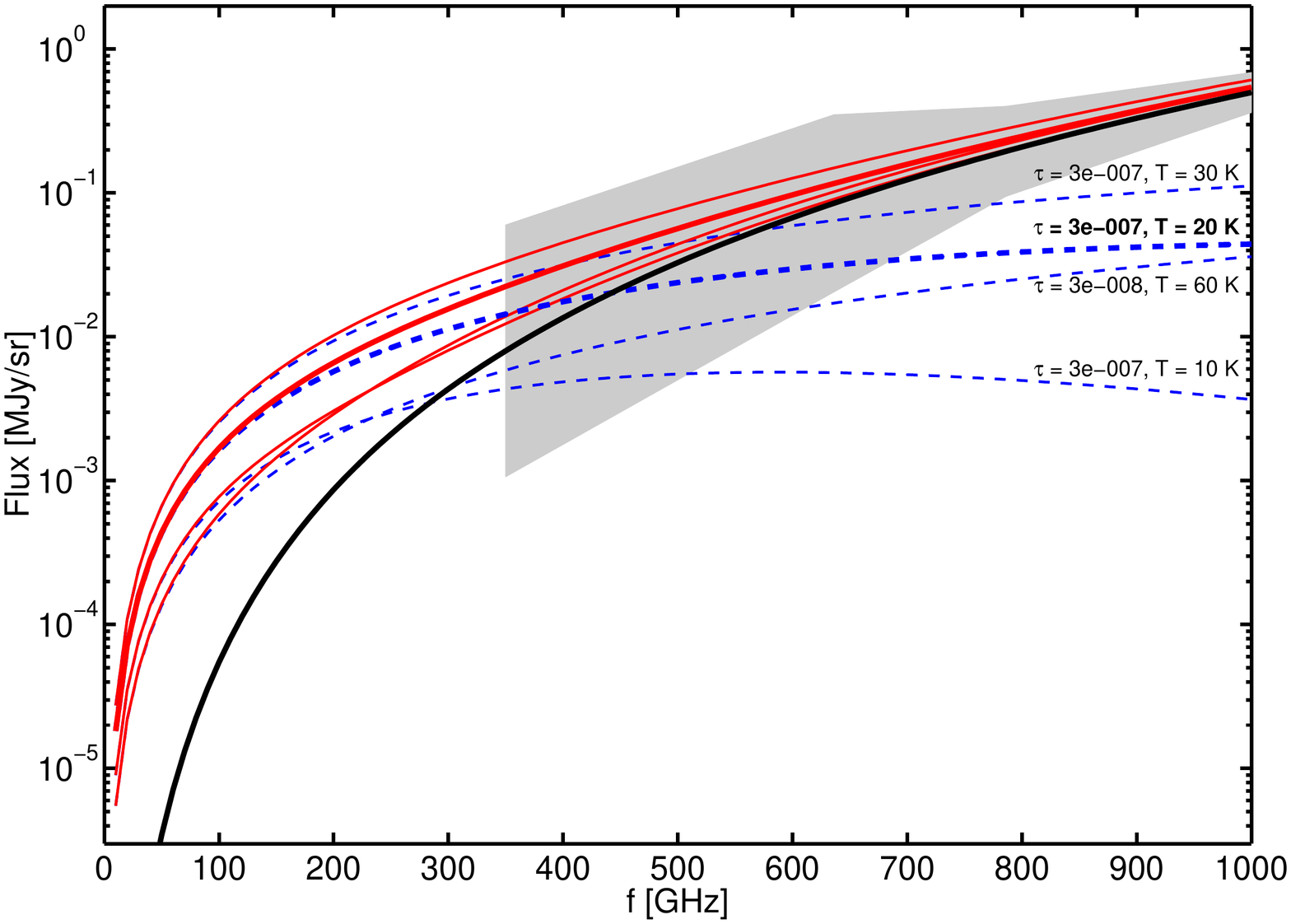}}
\end{tabular}
\caption{\label{fig:one}
Panel a):
 the ILC map derived from WMAP 7 yr data. We displaied also with black lines the ecliptic 
plane 
and the three increasing regions where the KBOE is present 
in the case of  our toy model, for three different increasing values of the height scale, 
$\Hwedge$ = $17.5\degr$, $35\degr$ and $70\degr$ 
Note that this figure has been not included in the pubblished version for 
editorial reasons.
Panel b):
Comparison of ZLE fluxes  compatible with  COBE/FIRAS data 
%\citep{Fixsen:Dwek:2002} 
(Fixsen \& Dwek 2002)
and a set of possible models of KBOE 
%\citep{Stern:1996}. 
(Stern 1996).
The black solid line shows the ZLE derived from the best fit model to COBE/FIRAS data and extrapolated to lower frequencies.
 The gray band represents a sketch of the allowed region obtained from the error bars in 
%\cite{Fixsen:Dwek:2002}. 
Fixsen \& Dwek (2002).
The blue dashed lines display four different models of KBOE corresponding to different values of 
$\taudust $ and $\Tdust $. 
The resulting fluxes, sum of KBOE and ZLE, are represented by the red solid lines.
Note that the classical ZLE (estimated on the basis of COBE data)
is negligible in practice at WMAP frequencies, whereas KBOE might not be ignored.
}
\end{figure}

%\section{Toy model of KBOE}\label{sec:wedge:model}

To explore the effect of the KBOE on CMB maps we considered a toy model 
in which the emission is confined within a region of width $\Hwedge$,
symmetric with respect to the ecliptic plane, with constant brightness inside and zero outside.
Owing to the planar and cylindrical symmetry of the model,
all the coefficients $a_{\ell,m}$ 
for the multipole expansion in ecliptic coordinates of
the KBOE will be zero except for those with even $\ell$ and $m = 0$.
Then KBOE (and likewise ZLE) will just affect the
map components with even $\ell$.
By denoting with $\IKBOE(\Point)$,
the KBOE for unit solid angle, 
integrated along a given pointing direction $\Point$, 
notable relations can be derived
for the sky--averaged KBOE, $\XWmean$, its variance over the sky
and its quadrupole $\atzwedge$.
Since in the model $\IKBOE(\Point)$ is constant for $\Point$ within an angle
$\pm\Hwedge/2$ from the ecliptic,
defining $\mathcal{S} = \sin(\Hwedge/2)$, we have
$\XWmean =  \AW  \mathcal{S} \, $, 
    $\var(\IKBOE)  =  \AW^2 \mathcal{S} \left({1-\mathcal{S}}\right) \, $, and
\begin{equation}
          \label{eq:wedge:quadrupole}
          \atzwedge  =  -\AW \sqrt{5\pi} \mathcal{S} \left(1-\mathcal{S}^2\right)       \; ;  
\end{equation}

\noindent
here $\AW$ is the constant value of KBO dust emission within the region.
The maximum variance is $\var(X) \le \AW^2/4$ and occurs for 
$\Hwedge = 60\deg$, while
the $|a_{2,0}|$ maximum is 
$|a_{2,0}| \le \sqrt{20\pi/27} \AW \approx 1.53 \AW$ occurring for 
$\Hwedge\approx70.53\deg$.
Flux variations can be translated in term of brightness temperature variations with the usual conversion factor
$%{\partial \Bcmb}/{\partial T} = 
\left[{\partial B_\nu(T)}/{\partial T}
\right]_{T=2.725 {\rm K}}$.
%  \begin{equation}
%  \frac{\partial \Bcmb}{\partial T} = \left(\frac{\partial B_\nu(T)}{\partial T}
%  \right)_{T=2.725^\circ\mathrm{K}}.
%  \end{equation}
%
%

\begin{table}
\begin{center}
 \begin{tabular}{cccccc}
%\hline
\hline
      &            & \multicolumn{3}{c}{WMAP Bands}\\
Model  &  $\Hwedge$ & Q               & V               & W \\
\hline
 CM: & $17.5\degr$  & 2.905       & 3.016       & 3.362 \\
 WM: & $17.5\degr$ & 1.211       & 1.223       & 1.256 \\
\hline
 CM: & $35\degr$           & 2.698       & 2.698       & 3.100 \\
 WM: & $35\degr$           & 1.194       & 1.204       & 1.234 \\
\hline
 CM: & $70\degr$           & 2.064       & 2.122       & 2.303 \\
 WM: & $70\degr$           & 1.135       & 1.142       & 1.162 \\
%\hline
\hline
\end{tabular}
\end{center}
\caption{
\label{tab:one}
Ratio between the quadrupole derived from the subtraction of 
the KBOE template from 
the WMAP 7~yr ILC map 
and the original WMAP map for the three WMAP Q, V, W bands
and the considered models, using {\tt anafast}.
Models are the Cold Model (CM), with $T=30$~K, $\tau=3 \times 10^{-7}$
and the 
Warm Model (WM) with $T=60$~K, $\tau=3 \times 10^{-8}$.
}
\end{table}

Detailed information about
the distribution of dust in the outer Solar System
to fix $\Hwedge$ and $\AW$ is missing.
A constraint on
$\Hwedge$ comes out from the distribution of inclinations of KBO orbits with respect
to the ecliptic. KBOs are divided in two populations:
classical KBOs with orbital inclinations within $10\degr$ about the ecliptic,
and scattered KBOs with orbital inclinations within $40\degr$ about the ecliptic. 
In analogy with the distribution of dust from the erosion of
Main Belt asteroids, we assumed that KBOs are a tracer of KBO dust and then of 
KBOE. Thus,
we selected three possible representative values of $\Hwedge = 17.5\degr$,
$35\degr$ and $70\degr$.
% DECOMMENTARE
The region of the ILC map affected by the KBOE in these three cases is shown in 
Fig.~\ref{fig:one}a 
(Note that this figure has been not included in the pubblished version for 
editorial reasons.).

To constrain $\AW$ is a more complicated task, since it depends on details about grains
such as
their temperature, their size and their radial distribution, 
as well as their shape and their mineralogical composition. 
We rely on existing models
predicting the possible  sky--averaged KBOE brightness,
leaving the development of a self--consistent model
as a topic for a subsequent work.
We choose as an example the model of \citet{Stern:1996}, based on the production of dust
in the KBO band by collisional disgregation of the KBOs. 
This model has the advantage to describe the sky--averaged KBOE in terms of 
two free parameters: 
the averaged dust temperature, $\Tdust$,
ranging from about 10~K to 60~K and the sky--averaged optical depth, $\taudust < 10^{-6}$.
The model has been already constrained by the authors using IR data. 
Constrains come also from the microwave
COBE/FIRAS data for the sky--averaged ZLE, $\IZLE$, \citep{Fixsen:Dwek:2002}.
%
%Fig.~\ref{fig:one}b 
Fig.~\ref{fig:one}b
compares the spectral energy distribution (SED) of the sky--averaged flux from KBOE, 
$\IKBOE$, 
from the 
\cite{Stern:1996} prescription with $\IZLE$.
All those
combinations of $(\Tdust, \taudust)$ values for which $\ITOT$ is upper-bounded 
by the gray band in 
%Fig.~\ref{fig:one}b 
Fig.~\ref{fig:one}b
are acceptable.
This is equivalent to take the $(\IKBOE/\IZLE)$ ratio
at 350~GHz to be less than $\simeq 6.5$.
%
%Fig.~\ref{fig:one}b  
Fig.~\ref{fig:one}b
shows that models with dust temperature $T = 10$~K, 30~K and 
60~K and $\tau\le 3\times10^{-7}$ are allowed by COBE/FIRAS data, and they
predict
at 93.5~GHz 
$\IKBOE$ in excess of up to $\simeq 80$ times $\IZLE$.
At frequencies above 1 THz, KBOE is one to three order of magnitudes 
below the classical ZLE,
and it does not display any seasonal effect, so that its contribution to 
the bulk ZLE emission seen in COBE/DIRBE is negligible.
Note that in the case of the most extreme models allowed by the COBE/FIRAS data
it is possible to predict a contribution to the quadrupole 
(see Eq.~(\ref{eq:wedge:quadrupole}))
of up to $\approx 10^2 \; \microK^2$, 
%which must be damped 
%by the geometrical factors in Eq.~(\ref{eq:wedge:quadrupole}),
to be compared with the expected ZLE contribution
less than $0.02 \; \microK^2$ and the CMB quadrupole moment 
$\Delta T_{\mathrm{CMB},\ell = 2} \sim 200 \; \microK^2$ 
(\cite{WMAP Explanatory Supplement}, see also footnote 1).

\begin{figure}
\centering
{\includegraphics[width=120mm,angle=0]{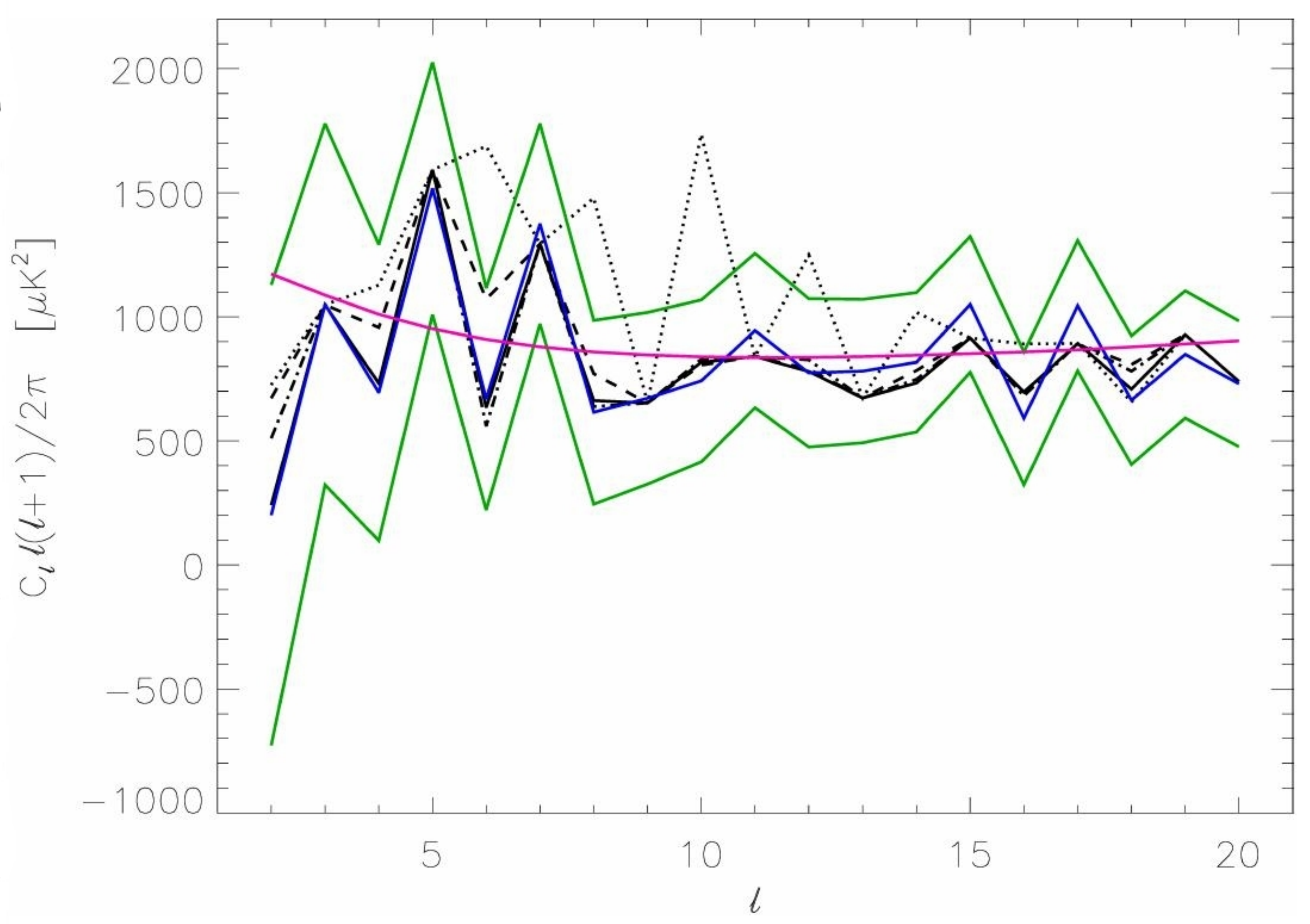}} 
\caption{
\label{fig:two:a}
APS at low multipoles as derived by the WMAP team analyzing the
ILC 7~yr map (blue solid line)
with its $1\sigma$ errors (green solid lines);
APS of the ILC 7~yr map derived using {\tt Anafast} 
(black solid line); 
APS of the ILC 7~yr map after the subtraction of our
KBOE template with $\Hwedge$ = $70\degr$, $35\degr$ and $17.5\degr$
(black dotted--dashed line, dashed line, and
dotted line, respectively).
The dust model with $\tau=3 \times 10^{-7}$ and $T=30$~K (CM) and the V band is here 
considered. The red solid line shows APS of the best fit 
$\Lambda\mathrm{CDM}$ model for the WMAP 7~yr map.
}
\end{figure}

How massive must be the debris disk to produce the KBOE?
An estimate of the required Kuiper Belt mass to be 
converted in dust, $\Mdust$, 
can be readily derived from $\tau$.
From dynamical models of dust propagation and the distribution of objects in the 
Kuiper Belt it is possible to infer that Kuiper Belt Dust should be confined within
$\Rin \approx 30$~AU and 
$\Rout \approx 50$~AU. Since $\Tdust \propto 1/\sqrt{R}$ 
the grains could be considered approximately isothermal. 
As a further approximation it is possible to assume 
the numerical density of the dust $\ndust$ to be uniform between $\Rin$ and
$\Rout$.
Assuming grains as spheres, with a typical diameter, $\dustsize$, 
unit emissivity efficiency, scattering cross--section given by
their geometrical cross section, then 
$\ndust \approx 4 \tau /[\pi \dustsize^2(\Rout - \Rin)]$ and 
$\Mdust \approx (8\pi/9)\tau \rho \dustsize \sin(\Hwedge/2)  [\Rout^3 - \Rin^3] / [\Rout - \Rin]$.
% \begin{equation}\label{eq:mass:dust}
%\Mdust \approx \frac{8\pi}{9} \frac{\Rout^3 - \Rin^3}{\Rout - \Rin} 
%\sin\left(\frac{\Hwedge}{2}\right) 
% \tau \rho \dustsize .
% \end{equation}
%
% \noindent
Fragments should be composed largely of silicates, ices, and carbon compounds
with densities $\rho$  ranging between 
1 and $3\;\mathrm{gr}/\mathrm{cm}^3$.
For $\tau \la 3 \times 10^{-7}$, $\Hwedge \la 70\degr$,
and
$\dustsize\approx1$~cm, $\Mdust$  varies between
$6\times 10^{23}$~gr and 
$18 \times 10^{23}$~gr,
i.e.
less than $0.14$ Pluto masses. 

\noindent
%
% Leading the dust temperatures T, to vary of about a 30\%, to be compared to the 
% factor of two variation expected for near grains in the MBA.

%
\section{Analysis}\label{sec:analysis}
The toy model outlined in 
%Sect. \ref{sec:wedge:model} 
Sect.~\ref{sec:introduction} 
is exploited to assess its possible implications
for CMB anisotropy statistical estimators and thus for cosmological models and parameters,
under the assumption that such a kind of component
present in the microwave maps, but it 
is hidden in the currently available CMB component maps. 

%\subsection{Template production}\label{sec:template}
\noindent
{\bf Template production --}
%
%\noindent
According to the above assumption, we subtract our toy models from the available CMB anisotropy map.
We will present here results obtained exploiting some 
%representative 
KBOE toy model templates 
characterized by $\Hwedge$ = $17.5\degr$, $35\degr$, and $70\degr$ for two combinations of averaged dust temperature
and sky-averaged optical depth: $\tau=3 \times 10^{-7}$, $T=30$~K 
(cold model, CM) 
and $\tau=3 \times 10^{-8}$, $T=60$~K 
(warm model, WM). 
These templates are computed in the ecliptic frame at the centers of the WMAP Q, V, and W frequency bands and generated at a resolution 
defined by the HEALPix 
\citep{Gorsky:etal:2005} 
\footnote{http://healpix.jpl.nasa.gov/}
parameter $N_{side}=256$ (i.e. with a pixel size of about $13.7'$), appropriate to our large scale analysis.
They are then convolved with a Gaussian symmetric beam with FWHM of $1\degr$ to match the beam resolution of the ILC 7~yr map 
released by the WMAP team and then transformed 
%from the ecliptic to 
into 
%the 
Galactic frame.
We have then subtracted the proper monopole to each template.
At each frequency, we subtract our KBOE templates from the WMAP ILC 7~yr map (previously degraded at  $N_{side}=256$)
to produce new CMB anisotropy maps cleaned by the adopted KBOE contribution.

\begin{figure}
\centering
\includegraphics[width=120mm]{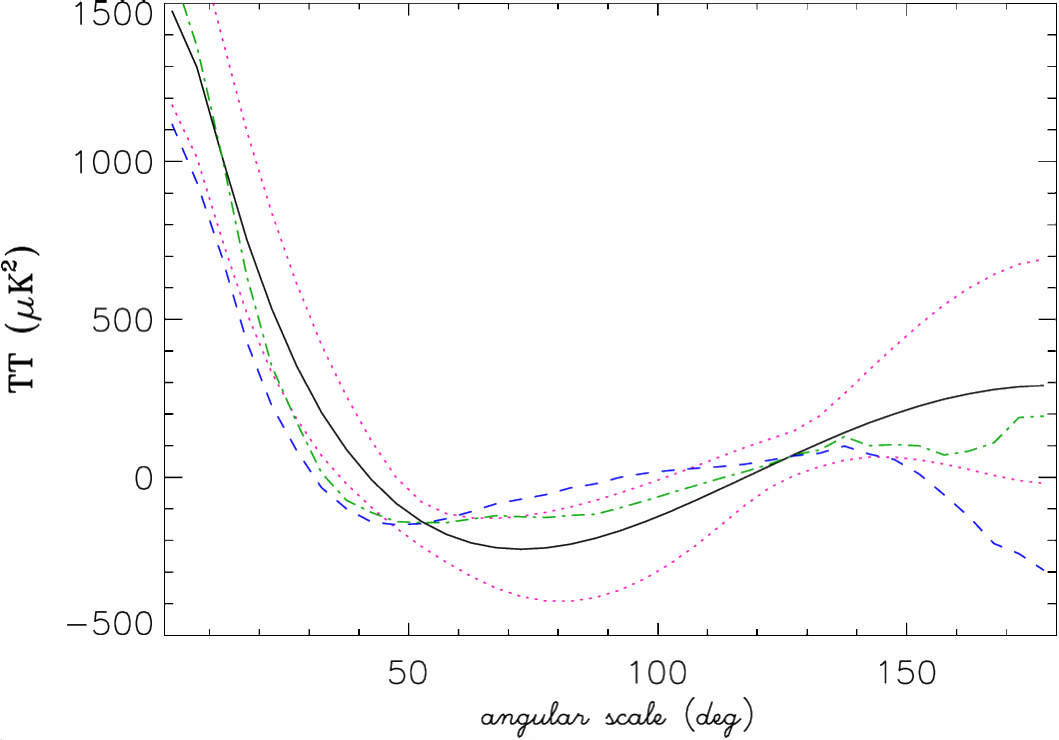} 
\caption{
\label{fig:two:b}
Two-points correlation function computed for maps at HEALPix resolution $\Nside = 16$.
The solid line displays the average of $10^5$ MC realizations of CMB anisotropy maps 
extracted from the WMAP 7~yr best fit $\Lambda\mathrm{CDM}$ model APS shown in 
  Fig.~\ref{fig:two:a}.
The red dotted lines show the corresponding $1 \sigma$ level fluctuations of the MC simulation.
The blue dashed line refers to the ILC 7~yr map. The green dotted-dashed line refers to the map derived 
subtracting from the ILC 7~yr map one of our KBOE model, namely that obtained for the CM with $\Hwedge$ = $17.5\degr$.
}
\end{figure}

%\subsection{Angular power spectrum}\label{sec:aps}
\noindent
{\bf Angular power spectrum --}
%
%\noindent
We extracted the whole--sky APS of the ILC 7~yr map 
released by the WMAP team and of the maps cleaned by the KBOE contribution described 
above using the publicly available HEALPix routine {\tt anafast}, applying 3 iterations and subtracting monopole and dipole. 
The APS derived for ILC 7~yr map is compared with that released by the WMAP team. The results are shown in 
Table~\ref{tab:one} and 
  Fig.~\ref{fig:two:a}
 for the low multipoles of interest here. As evident, no significant difference is found between 
the APS derived with  {\tt anafast} and with the finer approach applied by the WMAP team, thus probing that the results simply derived with 
{\tt anafast} are accurate enough for the aims of this work. 
Table~\ref{tab:one}  shows that the subtraction of the KBOE template from the ILC map in the case of the CM implies a larger 
enhancement of the CMB quadrupole, while the effect, although present, is not so remarkable in the case of the WM. 
Also, this enhancement increases with the decrease of $\Hwedge$.  
These results do not depend significantly on the considered 
frequency channel because of the considered KBOE spectral shape. 
Thus, we will focus in particular on the V band, where Galactic foregrounds are minimum at large scales, and on the CM.
In order to evaluate the probability that CMB quadrupole phases anti-correlate with the ``deterministic'' ones of KBOE,
we have performed a simple Monte-Carlo simulations. We extracted $10^4$ random quadrupole realizations from 
the best fit of $\Lambda$CDM model obtained by WMAP 7~yr data and, for each of them, we have added our template
and evaluated the modified quadrupole. 
We found that for about one out of three of realizations 
the band power, $\ell(\ell+1) C_\ell / (2\pi)$, of the quadrupole decreases from its primordial value, 
and, for example, for $\simeq 11.5 -13.4 \%$ of the events 
it is lowered more than 300 $\mu K^2$ depending on the specific CM considered \footnote{ Considering the fiducial 
$\Lambda$CDM model as a reference, the decreasing of at least  300 $\mu K^2$ occurs  for $\simeq 30 - 36\%$
 of the events.}.
These numbers do not depend strongly on the considered value of $\Hwedge$.
This shows, reversely, that the probability of observing an ILC map whose quadrupole turn to be increased, once the KBOE template is removed,
is not negligible.
  Fig.~\ref{fig:two:a}
  reports the APS up to $\ell = 20$. Note that only the power of even multipoles is significantly
affected, as expected since the geometrical symmetry of the KBOE component.
For $\Hwedge$ = $70\degr$ only the quadrupole turns to be significantly
amplified, while for $\Hwedge$ = $35\degr$ the power enhancement is important also at $\ell=4$ and 6 
and not negligible at $\ell=8$. Remarkably, for $\Hwedge$ = $17.5\degr$ the power increase is important 
for the even multipoles up to $\ell \simeq 16$ (see also 
%Sect. \ref{sec:parity}
the following discussion on Parity). 
Note that the overall effect, when KBOE template is subtracted from the ILC map, 
is an increase of the resulting CMB power at even multipoles. 
This is for geometrical reasons, but related to the relative phases,
or orientations, of $a_{\ell m}$ of ILC and of KBOE patterns.

\begin{figure}
\centering
\begin{tabular}{cc}
% 4.7,5
\includegraphics[height=7.52cm,width=8cm]{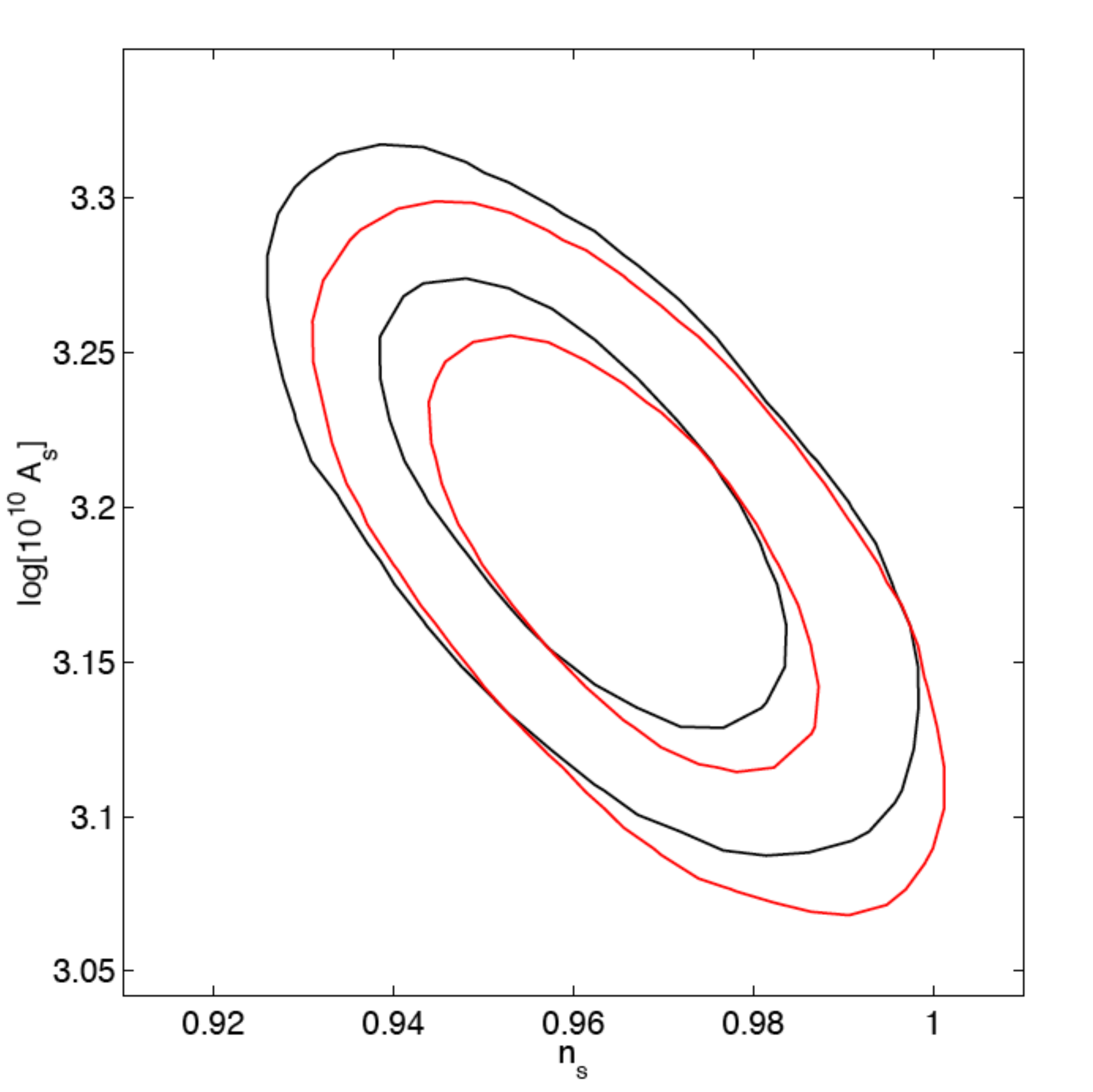} &
\includegraphics[height=7.52cm,width=8cm]{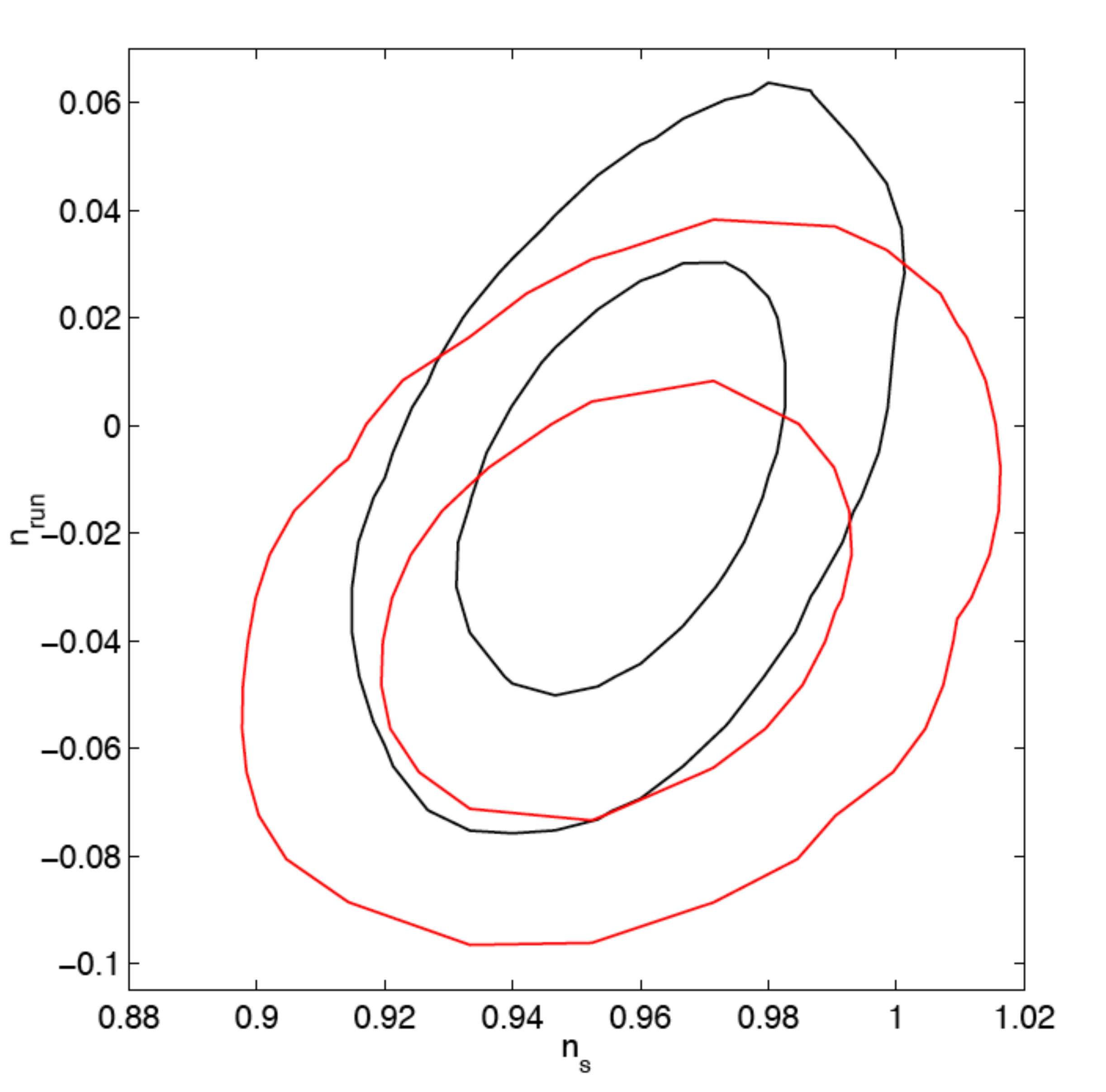} \\
\end{tabular}
\caption{
\label{fig:three}
Two dimensional marginalised 
probability distributions for cosmological parameters by removing (black lines) or not removing (red lines) 
the KBOE template for the CM with $\Hwedge$ = $35\degr$ 
(curves are the $68\%$ and $95\%$ confidence level). 
To the left (right) the plot for $A_s$ ($n_{\rm run}$) vs $n_s$ for a standard $\Lambda$CDM model (including running of the scalar 
spectral index).
}
\end{figure}

%\subsection{Correlation function}\label{sec:cf}
\noindent
{\bf Correlation function --}
%
%\noindent
The two point correlation function is defined as
%\begin{equation}\label{eq:two:point:correlation:function}
$C(\theta) = \langle T(\hat n_1) T(\hat n_2) \rangle _{\hat n_1 \cdot \hat n_2 = \cos (\theta)}
\, ,
$
%\end{equation}
%\noindent
where the symbol $\langle ... \rangle_{\hat n_1 \cdot \hat n_2 = \cos (\theta)}$ stands for the average over an ensemble of realizations
in which the directions $\hat n_1$ and $\hat n_2$ form an angle given by $\theta$.
In \cite{Copi:2008hw} 
it is shown that at angular scales greater than about $60\degr$ 
the WMAP data occur in $0.025$\% of realizations of the concordance model
(see also \cite{Copi:etal:2006,Hajian:2007pi,Copi:2010na} and references therein).
In 
  Fig.~\ref{fig:two:b}
we show $C(\theta)$ (black line) obtained through $10^5$ MonteCarlo (MC) random realizations
extracted from the WMAP 7~yr best fit $\Lambda$CDM model whose APS is displayed in
  Fig.~\ref{fig:two:a}.
These random realizations 
properly take into account the anisotropic white noise level of the $V$ band channel of WMAP 7~yr data.
The resolution we have considered for this analysis is given by the HEALPix parameter $N_{side}=16$ and 
the angle $\theta$ has been binned with a size of $5\degr$. 
The red dotted lines show the corresponding 1$\sigma$ level fluctuation of the MC simulation. The blue dashed line refers to the ILC 7~yr map. 
The green dotted-dashed line refers to the map derived subtracting from the ILC 7~yr map one of our KBOE model, 
namely that obtained for the CM with $H_{KBOE} = 17.5\degr$.
It is interesting to note how well the green line is (almost) always well consistent within 1$\sigma$ level with the $\Lambda$CDM model
whereas the blue line (i.e. the WMAP 7~yr ILC map) is often out of the 1$\sigma$ contour, especially at large angular scale.
This improvement is typical of our KBOE templates for the CM, while the templates for the WM do not change much the WMAP 7~yr ILC map correlation function.
% {\em \large MANCA LA CONCLUSIONE IN QUESTA SEZIONE NON SI CAPISCE COSA MIGLIORA
% FORSE SAREBBE PIU CHIARO SCRIVERE 
% {\tt IT IS EVIDENT HOW AFTER SUBTRACTING OUR KBOE MODEL FROM THE ILC MAP THE 
% $C(\theta)$ FOR THE RESIDUALS IS MORE SIMILAR TO THE EXPECTED $C(\theta)$ FROM 
% A PURE CMB PROCESS.
% }
% }

%\subsection{Cosmological parameters}\label{sec:parameters}
\noindent
{\bf Cosmological parameters --}
We quantify here the impact of the removal of the KBOE emission on
cosmological parameters.
The increase of the amplitude at low
multipoles has interesting consequences on some of the cosmological
parameters.
For this purpose, we use the WMAP 7~yr likelihood code publicly available 
with the option of a pixel likelihood code at low resolution and
substitute the WMAP 7~yr ILC map with the KBOE cleaned one.
The Markov Chain Monte Carlo package CosmoMC \citep{cosmomc} is connected with
this modified likelihood code.
As shown in 
  Fig.~\ref{fig:three}
the removal of a Solar System
contamination (namely the CM with $H_{KBOE} = 35\degr$)
modifies slightly the amplitude and the spectral index of scalar
perturbations (by choosing a pivot scale
$k_* = 0.002 {\rm Mpc}^{-1}$) in a standard
$\Lambda$CDM scenario characterized by $6$ cosmological parameters (plus
the amplitude of the residual Sunyaev-Zeldovich
effect): as expected by 
  Fig.~\ref{fig:two:a},
a  slightly larger value for
the amplitude of primordial fluctuations is preferred by
the larger amplitude of the temperature anisotropies at low $\ell$'s.
We find a little impact of the removal of this foreground on the 
constraints on the spatial flatness of our Universe. 
A larger effect is expected in extensions of the $\Lambda$CDM model in which
the small value of the quadrupole plays a relevant role: 
  Fig.~\ref{fig:three}
shows how
smaller values of the running of the scalar spectral index $n_{\rm run} =
d n_s/(d \ln k)$ are preferred once the KBOE
is removed (in this last case we have adopted an optimal pivot
scale $k_* = 0.017 {\rm Mpc}^{-1}$).
It is important to note how the change in the mean value of $n_{\rm run}$
we obtain is $0.023$ after removing the KBOE template.
Such difference is comparable
with the {\Planck} 
sensitivity to $n_{\rm run}$: it is therefore important to
assess foreground contamination on large scales
to infer on the primordial spectrum of density perturbations.
Also, since correcting for KBOE the low multipole tail in TT is not so suppressed,
including tensor perturbations in the analysis, we find that, the upper limit on tensor-to-scalar ratio of primordial perturbations, $r_{TS}$,  
is relaxed to $r_{TS} < 0.42$  (at 95\%CL), compared 0.36 given in  \cite{Larson:2010gs}.
Note that this effect is due to the larger relevance of temperature information with respect to 
polarization in WMAP data.

\begin{figure}
\centering
\includegraphics[width=120mm]{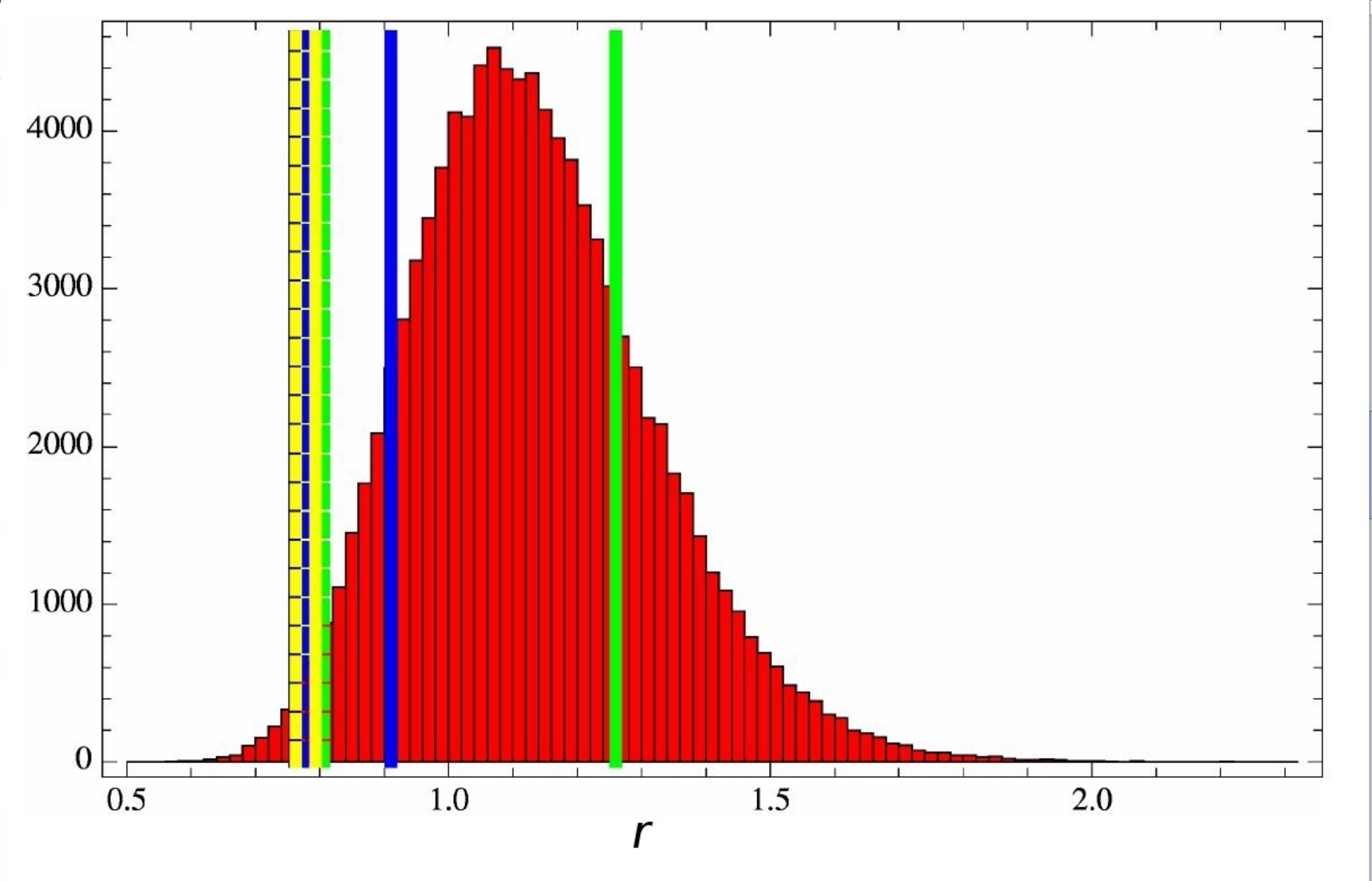} 
\caption{
\label{fig:four}
Parity anomaly of the estimator $r= P_+/P_-$ as defined in the text 
%(see Eq. (\ref{paritydef})
with $\ell_{max} = 22$. The histogram (in red) displays the distribution of $r$ obtained from 
$10^5$ MC realizations. Vertical lines correspond to the maps considered in this work: 
the black solid line (on the left) refers to the ILC 7~yr map;
colored solid lines refer to the CM; colored dashed lines refer to the WM.
Green, blue, and yellow lines are for $\Hwedge$ = $17.5\degr$, $35\degr$, and $70\degr$, 
respectively.
}
\end{figure}

%\subsection{Parity}\label{sec:parity}
\noindent
{\bf Parity --}
%
%\noindent
One of the large scale anomalies is the Parity anomaly. For all-sky maps it is customary to expand the CMB temperature fluctuations $T(\hat n)$
in terms of Spherical Harmonics $Y_{\ell m}(\hat n)$ where $\hat n$ is a direction in the sky,
namely depending on a couple of angles $(\theta, \phi)$, 
%\begin{equation}
$T(\hat n) = \sum_{\ell m} \, a_{\ell m} \, Y_{\ell m}(\hat n) \equiv  \sum_{\ell} T_{\ell}(\hat n)
\, ,
$
%\end{equation}
where in the last equation we have implicitly defined the maps $T_{\ell}(\hat n)$ for each angular scale $\ell$. 
Since under reflection (or Parity) symmetry $\hat n \rightarrow -\hat n$ the Spherical Harmonics
behave as
%\begin{equation}
$Y_{\ell m}(\hat n) \rightarrow (-1)^{\ell} \,Y_{\ell m}(\hat n)
\, ,
$
%\end{equation}
and the map $T_{\ell}(\hat n)$ has even parity for even $\ell$ and odd parity 
for odd $\ell$.

At large scales, we expect a Sachs-Wolfe plateau in terms of TT APS,
therefore we roughly forecast the same amount of power in even and odd maps. 
It has been recently proved that for  WMAP data at the large scales
\citep{Kim:2010gd,Kim:2010gf}, 
precisely in the range of $\ell = 2-22$, the total power coming from even $\ell$'s is 
unlikely smaller than the total power present in the odd $\ell$'s.
The probability related to this event is as low as $0.4\%$ for the WMAP 7~yr data \cite{Kim:2010gf}
(see \citet{Gruppuso:2010nd} for a temperature and polarization joint analysis).
The analysis of this power asymmetry was first proposed in \cite{Land:2005jq} 
to test the presence of foreground residuals since templates
of dust, free-free or synchrotron emission possess a large Parity asymmetry.
Also the KBOE template we propose in this paper, as well as the standard ZLE, is highly Parity-asymmetric.
Therefore it is worth to study its possible impact on the WMAP ILC 7~yr map, of course under the assumption that a residual 
of the kind considered in the current paper, is present in the ILC map.

We consider the same estimator of \cite{Kim:2010gd,Kim:2010gf} defined as
$r=P^{+}/P^{-}$, where
%\begin{equation}\label{eq:parity}
$P^{+/-} = \sum_{\ell^{+/-}} \ell (\ell +1) \, C_{\ell}/ 2 \pi \, ;$
%\label{paritydef}
%\end{equation}
%\noindent
here with $\sum_{\ell^{+/-}}$ we mean the sum over even or odd $\ell$'s respectively
in the considered range (here $2-22$ \footnote{We consider the same range of \cite{Kim:2010gd}
where it was shown the maximum of the anomaly. See also \cite{Bennett:2010jb} for comments
about ``a posteriori'' selections.}).
We have extracted $10^5$ random maps from the best fit model of WMAP 7~yr working at 
the HEALPix resolution $N_{side}=64$
(i.e. all-sky maps of $49152$ pixels). We have computed the $r$ estimator for each random map 
and we have built the probability distribution function (pdf) for $r$, shown in
  Fig.~\ref{fig:four}.
Note that the pdf for $r$ does not peak around the value 1, but slightly larger since for the chosen $\ell$-range,
there are more terms at numerator than in the denominator. Vertical lines in
  Fig.~\ref{fig:four}.
represent the values of $r$
for the considered maps. Black vertical line (on left) stands for the WMAP 7~yr ILC map. 
The probability to have this value is as low as $0.91\%$ \footnote{This value is slightly larger than what quoted in \cite{Kim:2010gd} that is $0.86\%$.
This might be due to the details of the MC, like number of simulations, resolutions of maps and the considered FWHM.}.
When we remove our KBOE template from the maps we have the colored vertical lines.
Green, blue, and yellow lines stand for $H_{KBOE} = 17.5\degr, 35\degr, $ and $70\degr$, respectively.
Colored solid lines refer to the CM, colored dashed lines refer to the WM. 
Note that the probability associated to the green line (i.e. CM and $H_{KBOE} = 17.5\degr$) is $77.1\%$ and
the probability associated to the blue line (i.e. CM and $H_{KBOE} = 35\degr$) is $10.6\%$ .
The Parity anomaly is removed when these templates are properly taken into account.
In the definition of the Parity estimator in the current paper we adopted
$\ell_{max}=22$, just for simplicity and since for that $\ell_{max}$ value the anomaly is
remarkable. However, as shown in \cite{Kim:2010gd,Kim:2010gf,Gruppuso:2010nd},
in the WMAP TT spectrum there is a whole multipole range, 
%($\ell_{max}$ between about 15 and 25), 
rather than a single $\ell_{max}$ value, where the WMAP 7 parity
anomaly holds. This dims significantly the case for posterior biasing.
{
In analogy to what performed for the quadrupole, 
we have implemented a simple Monte-Carlo simulation
in order to evaluate the probability that CMB even multipoles phases (up to $\ell_{max}=22$) 
anti-correlate with the ``deterministic'' ones of KBOE.
We extracted $10^4$ random CMB realizations from the best fit of $\Lambda$CDM model obtained by WMAP 7~yr data and, 
for each of them, we have added our template and evaluated the modified spectrum. 
Hence, considering only the CM,  we have found that the probability of lowering the estimator $r$ is
$24.7\%$, $7.05\%$ and  $0.07\%$ for $\Hwedge=70^{\circ}$, $35^{\circ}$ and 
$17.5^{\circ}$, respectively.
Similarly, we computed the probability of lowering the ``even powers'' up to $\ell_{max}=22$ for some fixed threshold.
For instance, for $\Hwedge=70^{\circ}$ the probability to decrease the total even band powers of at least $300 \mu K^2$ with respect to the realization 
(or to the fiducial $\Lambda$CDM model) is $7.46\%$ $(43.46\%)$, whereas it is $2.67\%$ $(23.26\%)$ for  $\Hwedge=35^{\circ}$  and $0.01\%$ $(0.67\%)$ for $\Hwedge=17.5^{\circ}$. 
This shows that for $\Hwedge$ larger than few tens of degrees 
the anti-correlation probability turns out to be not negligible.
These numbers show that, contrarily to the quadrupole case discussed above, in this case the dependence on $\Hwedge$ is important.}  

\begin{figure}
\centering
% [height=48.7mm,width=40mm]
\includegraphics[width=120mm]{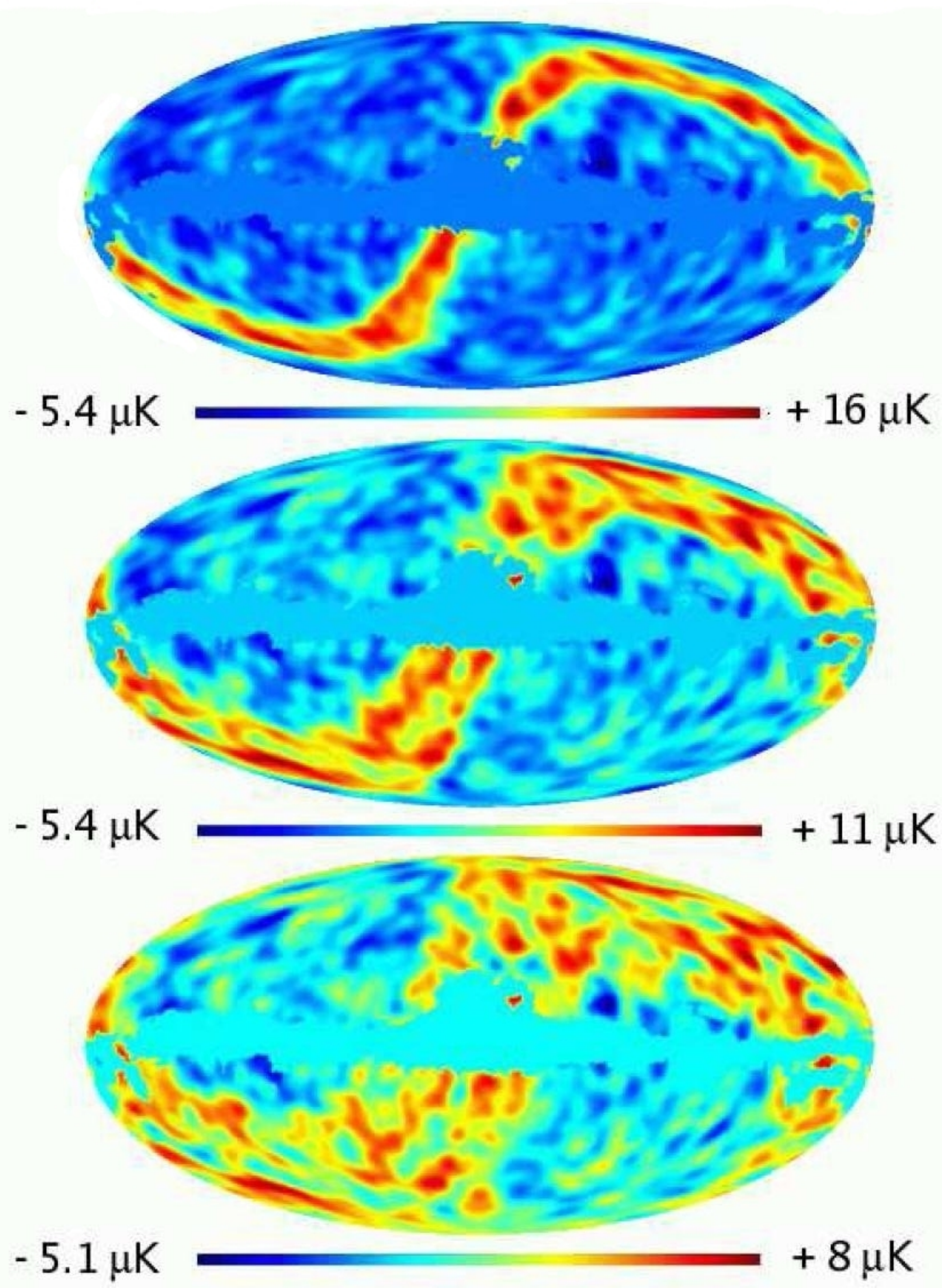} 
\caption{
\label{fig:five}
Predicted signal in the combination $V+W-2Q$ from the CM with $H_{KBOE} = 70^{\circ}$ (bottom), 
$H_{KBOE} = 35^{\circ}$ (middle), and $H_{KBOE} = 17^{\circ}$ (top). 
The map units are $\mu$K and it has been smoothed with a $7^{\circ}$ Gaussian. 
Compare this plot with Fig.~1 of 
%\cite{Diego:etal:2009} 
Diego et~al. (2010)
based on WMAP data:
the model with $H_{KBOE} = 70^{\circ}$ seems to agree better with the observations. 
}
\end{figure}

%\subsection{Alignments}\label{sec:alignement}
\noindent
{\bf Alignments --}
%
%\noindent
We have seen how the removal of the KBOE template (for the CM) can 
reduce the unlikeliness of some estimators, as the $\Cl$ and those
related to them.
Our preliminary tests carried out using 
a public code for the multipole vectors decomposition 
\footnote{See also http://www.phys.cwru.edu/projects/mpvectors/.}
by \citet{Copi:etal:2004} show that this does not happen for the estimators based 
on $a_{\ell m}$ phases by means of which alignment anomalies are detected.

%\subsection{Ecliptic excess in WMAP data}\label{sec:spot}
\noindent
{\bf Ecliptic excess in WMAP data --}
%
%\noindent
In \cite{Diego:etal:2009}, the authors look at the linear combination $V+W-2Q$ of 
WMAP bands. In their analysis, the WMAP bands have been cleaned of Galactic contamination 
and the bright point sources have been masked. The linear combination $V+W-2Q$ gets 
rid off the CMB signal completely and should contain a linear combination of the 
instrumental noise in each band, beam effects and residual Galactic and extragalactic 
foregrounds not accounted for in the cleaning (and masking) process. After masking 
out the Galactic plane, the authors found a significant large scale signal aligned 
with the ecliptic plane with a grey-body spectrum (see Fig.~1 in \cite{Diego:etal:2009}).
A Galactic origin for this signal was ruled out due to the high latitudes where 
it was found. An extragalactic origin was also ruled out due the highly anisotropic 
distribution of this signal. In that paper it was suggested a possible origin in the ZLE 
but extrapolations from actual measurements from COBE/FIRAS predicted 
a much smaller signal than needed to account for the excess in the ecliptic plane. 
The present paper offers a fresh alternative to explain the excess by assuming 
a different population of larger dust grains. We have combined the signal expected
from our KBOE models with the instrumental noise and beam of WMAP and built templates of the 
$V+W-2Q$ maps. In 
  Fig.~\ref{fig:five}
we show the predicted signal from the three CM models. 
The comparison with the signal detected by  \cite{Diego:etal:2009} (see their Fig.~1) shows 
that the model with $H_{KBOE} = 70^{\circ}$ seems to agree better with the observations than the others. 
The areas near the Galactic plane show no evidence of excess appearing in WMAP data, that could be due to residual synchrotron
emission.
%that would reduce the strength of the KBOE signal.
Note that while KBOE should be positive in  $V+W-2Q$ , 
the synchrotron signal shows up as a negative signal in the combination $V+W-2Q$.

\section{Conclusion}\label{sec:conclusions}

We exploited a simple toy model for diffuse emission 
from cold and large dust grains 
expected to exist in the outer part of the Solar System and 
requiring only a modest amount of mass,
with signal amplitudes, constrained in the Far--IR
by COBE data, compatible with simulations existing in literature.
We have produced and analyzed templates derived subtracting our toy model  
from WMAP ILC 7~yr maps to investigate on cosmological implications of such a foreground.
We find that the anomalies related to the low quadrupole of the angular 
power spectrum, the two point correlation function, and the Parity are significantly alleviated.
No significant impact has been found on low multipole alignments; of course, 
this does not exclude that more detailed models constructed in this framework could have consequences for these estimators.

It is interesting to note that there might be a relationship between the lack 
of effect of our model on the alignment of the low multipoles and the low amplitude 
of the quadrupole measured by WMAP. If the CMB intrinsic quadrupole 
is anti-correlated (i.e. with a similar pattern but opposite sign) 
with the ecliptic plane, any signal originating in our 
Solar System (like those considered in this paper) would {\it compete} with 
the CMB quadrupole, reducing its measured amplitude. On the other hand, 
the subtraction of the Solar System signal to the measured quadrupole would not 
change the orientation of the resulting signal since the subtraction operation 
transforms the anti-correlated signal in a correlated one and both patterns add up. 
A visual example of this 
%mechanism 
can be found for instance in Fig.~9 of
\cite{Diego:etal:2009} where also a hypothetical signal that traces 
the ecliptic plane is assumed. The same authors also demonstrate how signals in the ecliptic 
plane affect (significantly) only the even multipoles. Consequently, the octupole 
($\ell=3$) would remain basically unchanged.
The KBOE is able to explain the excess of signal found in \cite{Diego:etal:2009}, 
in particular for cold dust particles and for a high $H_{KBOE}$.

Finally, we showed that this foreground has an impact for some cosmological parameters characterizing 
the spectrum of primordial density perturbations, relevant for on-going and future CMB anisotropy experiments.  

Clearly, the model needs to be improved in the future and tested against the data 
from the {\Planck} satellite. Our analysis shows that it will be relevant to include, or at least to test, 
such component in the analysis of microwave anisotropy data at large scales.

\section*{Acknowledgements}
We acknowledge the use of the Legacy Archive for Microwave Background Data Analysis (LAMBDA)
%. 
%Support for LAMBDA is provided 
supported by the NASA Office of Space Science. 
Some of the results in this paper have been derived using the HEALPix \citep{Gorsky:etal:2005} package. 
We acknowledge the use of the public code for the multipole vectors decomposition
by \citet{Copi:etal:2004}.
Some of the simulations presented in this work have been performed 
using the computational facility of IASF Bologna and SP6 at CINECA.
We acknowledge partial support 
%for this work 
by the ASI/INAF Agreement 
I/072/09/0 for the {\Planck} LFI Activity of Phase E2 and the ASI contract I/016/07/0 COFIS.
M.M. 
%kindly 
acknowledges partial support by FFO Ricerca Libera 2009 and 2010. 
C.B. warmly thanks P. Naselsky and D.J. Schwarz for stimulating conversations. 
JMD acknowledges support from the proyect AYA2010-21766-C03-01.

\bsp

%\input bibl3.tex

%
% \cite{ref} per una citazione tra parentesi
% \cite[see]{ref} per dire "see ref"
% \citeasnoun{ref} per una citazione non tra parentesi
% \cite{ref1,ref2} per citazioni multiple (niente spazi)

% bibligraphy
%
% \clearpage

% \clearpage

\end{document}